%% file: draft-gravitino_vf.tex
\documentclass[12pt]{article}

\usepackage[utf8]{inputenc}
\usepackage{cancel}
\usepackage{amssymb}
\usepackage{soul}
\usepackage{tabularx}
\usepackage{xcolor}
\usepackage{booktabs}
\usepackage{subcaption} 
\usepackage{colortbl}
\usepackage{authblk}
\DeclareUnicodeCharacter{00A0}{ }

\newcolumntype{C}{>{\centering\arraybackslash}X}
\input{defsV3}
\allowdisplaybreaks

\usepackage{array, makecell}
\usepackage{boldline}
\usepackage{cite}

\title{\bf{MeV-GeV~$\gamma$-ray~telescopes~probing gravitino~LSP~with~coexisting~axino~NLSP~as dark~matter~in~the~$\mu\nu$SSM
}}
\author[a]{Germán A. G\'omez-Vargas
\thanks{{\it `Currently, data scientist corporativo at Derco'}}
\thanks{germangomez@derco.cl}}
\author[b,c]{Daniel~E.~L\'opez-Fogliani\thanks{daniel.lopez@df.uba.ar}}
\author[d,e]{Carlos~Mu\~noz\thanks{c.munoz@uam.es}} 
\author[b]{Andres~D.~Perez\thanks{andres.perez@df.uba.ar}}
\affil[a]{Instituto de Astrof\'isica, Pontificia Universidad Cat\'olica de Chile, 
Avenida Vicu\~na Mackenna 4860 Santiago, Chile}
\affil[b]{Instituto de F\'isica de Buenos Aires UBA \& CONICET, Departamento de F\'isica,
Facultad de Ciencia Exactas y Naturales, Universidad de Buenos Aires, 
1428 Buenos Aires, Argentina}
\affil[c]{
{Pontificia Universidad Cat\'olica Argentina, 
1107 Buenos Aires, Argentina}}
\affil[d]{Departamento de F\'{\i}sica Te\'{o}rica, Universidad Aut\'{o}noma de Madrid (UAM),
Campus de Cantoblanco, 28049 Madrid, Spain}
\affil[e]{Instituto de F\'{\i}sica Te\'{o}rica (IFT) UAM-CSIC, 
Campus de Cantoblanco, 28049 Madrid, Spain}
\date{\today}

\begin{document}
\maketitle
\begin{abstract}

In $R$-parity violating supersymmetry, the gravitino as the lightest supersymmetric particle (LSP) is a good candidate for dark matter, with the interesting characteristic to be detectable through $\gamma$-ray telescopes. 
We extend this analysis considering an axino next-to-LSP (NLSP) as a coexisting dark matter particle contributing with a detectable signal in the $\gamma$-ray spectrum. 
The analysis is carried out in the framework of the $\mu\nu$SSM, which solves the $\mu$ problem reproducing simultaneously neutrino data only with the addition of right-handed neutrinos.
We find that important regions of the parameter space can be tested by future
MeV-GeV $\gamma$-ray telescopes through the line signal coming from the decay of the axino NLSP into photon-neutrino.
In a special region, a double-line signal from axino NLSP and gravitino LSP is possible with both contributions detectable.

\end{abstract}


\pagebreak

\tableofcontents 

\section{Introduction}
\label{sec:intro}

Gravitino ($\psi_{3/2}$) {as the lightest supersymmetric particle (LSP)} or axino ($\tilde a$) LSP are interesting candidates for dark matter (DM) in the framework of supersymmetry (SUSY). In addition, in SUSY models where 
there is $R$-parity violation (RPV) these particles decay with a lifetime longer than the age of the Universe, producing a line signal potentially detectable
 in $\gamma$-ray telescopes. 
This was analyzed for the 
gravitino LSP in Refs.~\cite{Takayama:2000uz,Buchmuller:2007ui,Bertone:2007aw,Ibarra:2007wg,Ishiwata:2008cu,Choi:2010xn,Choi:2010jt,Diaz:2011pc,Restrepo:2011rj,Kolda:2014ppa,Bomark:2014yja} 
in the context of bilinear/trilinear RPV models~\cite{Barbier:2004ez},
and in Refs.~\cite{Choi:2009ng,GomezVargas:2011ph,Albert:2014hwa,GomezVargas:2017} in the
`$\mu$ from $\nu$' supersymmetric standard 
model ($\mu\nu$SSM)~\cite{propuvSSM}.
Similar analyses for axino LSP in bilinear/trilinear RPV models were carried out 
in 
Refs.~\cite{Kim:2001sh,Hooper:2004qf,Chun:2006ss,Endo:2013si,Kong:2014gea,Choi:2014tva,Liew:2014gia,Colucci:2015rsa,Bae:2017tqn,Colucci:2018yaq}.  

In a recent work~\cite{GomezVargas:2019vci}, we analyzed a multicomponent DM scenario with axino LSP and gravitino {next-to-LSP (NLSP)}. This is a decaying dark matter (DDM) scenario, where the gravitino in addition to the RPV decay channel into photon-neutrino, also undergoes an $R$-parity conserving (RPC) decay into axion-axino. 
The analysis was carried out in the framework of the $\mn$, where couplings involving right-handed neutrinos are introduced solving the $\mu$-problem and reproducing simultaneously the neutrino data~\cite{propuvSSM,analisisparam,Ghosh:2008yh,neutrinocp,Ghosh:2010zi}. A brief discussion and bibliography about the interesting phenomenology associated to the $\mn$, can be found in 
Ref.~\cite{GomezVargas:2019vci}, where it was shown that
significant regions of the parameter space of the model can be probed through a line signal coming from the axino decay. A double-line signal as smoking gun through 
the further contribution of the gravitino decay is also possible in a subset of those regions.

Here we want to extend our previous work, analyzing the opposite DDM scenario where the gravitino is the LSP and the axino the NLSP. 
Their masses, although model dependent
can be of the same order {or several orders of magnitude different} in realistic 
scenarios~\cite{Kawasaki:2013ae,Goto:1991gq,Chun:1992zk,Chun:1995hc,Kim:2012bb} such as in supergravity. Therefore, if the gravitino is the LSP the axino becomes naturally the NLSP. As a consequence, the axino decays into the gravitino plus an axion.
In this not yet explored RPV scenario, we will study its cosmological properties as well as the associated $\gamma$-ray constraints on spectral lines coming from current detectors such as \Fermi LAT, and prospects for future $\gamma$-ray space missions such as e-ASTROGAM~\cite{eAstrogamAngelis:2017} and AMEGO~\cite{AmegoCaputo:2017}.
Let us finally remark that, unlike RPC models, in the presence of RPV there is no gravitino problem, since the heavier SUSY particles decay to standard model particles much earlier than BBN epoch via RPV interactions. Related to this, in the following we will assume that the properties of the bosonic superpartner of the axion, the saxion, are appropriate not to spoil BBN predictions (see e.g. Refs.~\cite{Co:2017orl,Hamaguchi:2017}).

The paper is organized as follows. 
In Section~\ref{sec:decays}, first we briefly review the simple scenario with gravitino as LSP. 
In second place, we discuss the multicomponent DDM scenario with the axino as the NLSP. 
We will show the axino NLSP decay rates into photon plus neutrino and into gravitino LSP plus axion, and its contribution to the relic density.
Then, we will compute the $\gamma$-ray flux produced in this scenario.
Armed with these results, 
in Section~\ref{sec: results} we will show exclusion limits and prospects for detection  under the assumption of decaying gravitino LSP plus axino NLSP. 
This scenario is fully explored along with its parameter space allowed by cosmological observations such as dark radiation constraints. Finally, we will present the $\gamma$-ray measurements by current detectors employed to probe RPV SUSY parameter space region, focusing on the $\mu\nu$SSM, and we will discuss the prospect for detection in the case of the proposed e-ASTROGAM instrument.
The conclusions are left for Section~\ref{sec:conclusions}.

\section{Gravitino LSP and axino NLSP as dark matter}
\label{sec:decays}

In the framework of supergravity, both gravitino and axino have in the Lagrangian an interaction term with {the} photon  and {the} photino. In the presence of RPV, {photinos} and left-handed neutrinos are mixed in the neutral fermion mass matrix, and therefore the gravitino LSP, as well as the axino NLSP, are able to decay into photon and neutrino through this interaction term. This has significant implications because the 
signals are sharp $\gamma$-ray lines with energies $m_{3/2}/2$ and $m_{\tilde{a}}/2$, that could be detected in $\gamma$-ray satellite experiments such as \Fermi LAT, or future MeV-GeV telescopes as the proposed e-ASTROGAM~\footnote{
The flux of monochromatic neutrinos in RPV models
could be in principle observed in neutrino 
detectors~\cite{Covi:2008jy}. In the energy range MeV$-$GeV of interest for this work, the best limits on
DM decay into neutrinos correspond to the Super-Kamiokande 
detector~\cite{FRANKIEWICZ:2018lsh}. However, it is
straightforward to check that they
are about five orders of magnitude weaker compared to searches with the same targets using $\gamma$-rays.
Thus we concentrate on $\gamma$-ray line searches throughout this work.}.
In addition, the axino NLSP can decay into gravitino LSP and axion.
We will study in the next subsections the implications of this scenario for DM and its detectability, reviewing first the gravitino LSP decay.

{Concerning the gravitino mass, let us point out that in supergravity models it is related to the mechanism of SUSY breaking. In particular, in 
gravity-mediated SUSY breaking models, where the soft scalar masses are typically determined by the gravitino mass, it is sensible to expect the latter in the range GeV-TeV~\cite{Brignole:1997dp}, i.e. around the electroweak scale. However, specific Kahler potentials and/or superpotentials of the supergravity theory could allow for different situations, producing gravitinos with masses several orders of magnitude smaller than the electroweak scale. This is e.g. the case of no-scale supergravity models, where the gravitino mass is decoupled from the rest of the SUSY particle spectrum, and hence is possible to assign for it a mass much smaller than the electroweak 
scale~\cite{Ellis:1984kd}. {Alternatively}, very small gravitino masses with respect to the electroweak scale are obtained in gauge-mediated SUSY breaking models~\cite{Giudice:1998bp}. Also, e.g. in F-theory GUTs with the latter SUSY breaking mechanism working, one can obtain a gravitino mass 
of about $10-100$ MeV~\cite{Heckman:2009mn}.
Given the model dependence of the gravitino mass, we consider appropriate for our phenomenological work below not to choose a specific underlying supergravity model, and treat the mass as a free parameter.}

\subsection{Gravitino LSP decay}
\label{sec:gravitino lifetimes}

Gravitino decay width into photon-neutrino through RPV couplings is given by \cite{Borgani:1996ag,Takayama:2000uz}:
\bea
\Gamma(\psi_{3/2}\rightarrow\gamma\nu_i)\simeq\frac{m_{3/2}^3}{32\pi M_{P}^2}|U_{\tilde{\gamma} \nu}|^2\ ,
\label{decay2bodygravitino}
\eea
where $\Gamma(\psi_{3/2}\rightarrow\gamma\nu_i)$ denotes a sum of the partial decay widths into {$\gamma\nu_i$ and $\gamma\overline{\nu}_i$}, 
$m_{3/2}$ is the gravitino mass, 
$M_P\approx 2.43 \times 10^{18}$ GeV is the reduced Planck mass, and
the mixing parameter $|U_{\tilde{\gamma} \nu}|$ determines the photino content of the neutrino, 
\bea
\left|U_{\tilde{\gamma} \nu}\right|^2= \sum^3_{i=1}\left|N_{i1} \, \cos\theta_W +  N_{i2} \, \sin\theta_W\right|^2.
\label{photino}
\eea
Here $N_{i1} (N_{i2})$ is the bino (wino) component of the $i$-th neutrino, and $\theta_{W}$ is the weak mixing angle.
As obtained in Refs.~\cite{Choi:2009ng,GomezVargas:2017}, performing scans in the low-energy parameters of the $\mn$ in order to reproduce the observed neutrino masses and mixing angles, natural values of $|U_{\tilde{\gamma} \nu}|$ are in the range 
\begin{equation}
10^{-8} \lesssim |U_{\widetilde{\gamma}\nu}| \lesssim 10^{-6}\ ,
\label{relaxing}
\end{equation}
although relaxing some of the assumptions such as an approximate GUT relation for gaugino masses and/or TeV scales, the lower bound can be smaller:
\begin{equation}
10^{-10} \lesssim |U_{\widetilde{\gamma}\nu}| \lesssim 10^{-6}\ .
\label{relaxing2}
\end{equation}

As we can see in Eq.~(\ref{decay2bodygravitino}), the gravitino decay is suppressed both by the small RPV mixing parameter $|U_{\widetilde{\gamma}\nu}|$, and by the scale of the gravitational interaction, making its lifetime much longer than the age of the Universe $\tau_{3/2}\gg t_{today}\sim 10^{17}$ s, with
\begin{equation}
{\tau}_{3/2}=\Gamma^{-1}(\psi_{3/2}\rightarrow\gamma\nu_i)
\simeq 3.8\times 10^{33}\, {s}
\left(\frac{10^{-8}}{|U_{\widetilde{\gamma}\nu}|}\right)^2
\left(\frac{0.1\, \mathrm{GeV}}{m_{3/2}}\right)^{3}\ .
\label{gravitinolifetime}
\end{equation}

\subsection{Axino NLSP decays}
\label{sec: axino NLSP}

Axino partial decay width into photon-neutrino through RPV couplings satisfies \cite{Covi:2009pq}:
\bea
\Gamma(\tilde{a}\rightarrow\gamma\nu_i)\simeq\frac{m_{\tilde{a}}^3}{128\pi^3 f_a^2}\alpha_{em}^2C_{a\gamma\gamma}^2|U_{\tilde{\gamma} \nu}|^2\ ,
\label{decay2bodyaxino}
\eea
where $\Gamma(\tilde{a}\rightarrow\gamma\nu_i)$ denotes a sum of the partial decay widths into {$\gamma\nu_i$ and $\gamma\overline{\nu}_i$}, 
$m_{\tilde a}$ is the axino mass, 
$C_{a\gamma\gamma}$ is a model dependent constant of order unity, 
$\alpha_{em}=e^2/4\pi$, and $f_a$ is the Peccei-Quinn (PQ) scale.
This is the dominant decay for an axino LSP in the context of the $\mn$~\cite{GomezVargas:2019vci}, 
and is suppressed both by the small RPV parameter $|U_{\tilde{\gamma} \nu}|$ and by 
the large PQ scale $f_a \gtrsim 10^9$ GeV as obtainted from the observation 
of SN1987A~\cite{Kawasaki:2013ae}. 
It is worth noticing here that in comparison with Eq.~(\ref{decay2bodygravitino}),
$\Gamma(\tilde{a}\rightarrow\gamma\nu_i) > \Gamma(\psi_{3/2}\rightarrow\gamma\nu_i)$ since $f_a<M_P$, and $m_{\tilde a}>m_{3/2}$ for the case we are interested in this work with axino NLSP and gravitino LSP.
We can also compare Eq.~(\ref{gravitinolifetime}) with
\begin{equation}
\Gamma^{-1} (\tilde{a}\rightarrow\gamma\nu_i) \simeq 3.8\times 10^{25}\, {s}
\left(\frac{f_a}{10^{13}\, \mathrm{GeV}}\right)^2
\left(\frac{10^{-8}}{|U_{\widetilde{\gamma}\nu}|}\right)^2
\left(\frac{1\, \mathrm{GeV}}{m_{\tilde{a}}}\right)^{3},
\label{axinolifetime}
\end{equation}
where to write this equation we have assumed $C_{a\gamma\gamma}=1$.

Since in the framework of supergravity the axino has an interaction term with gravitino and axion, we have also to consider this RPC partial decay width~\cite{Hamaguchi:2017}
\begin{equation}
\Gamma(\tilde{a} \rightarrow \psi_{3/2} \, a)\simeq\frac{m_{\tilde{a}}^5}{96\pi m_{3/2}^2 M_P^2}(1-r_{3/2})^2(1-r_{3/2}^{2})^3, 
\label{decayto32a}
\end{equation}
where the axion mass has been neglected, and
\bea
r_{3/2} \equiv \frac{m_{3/2}}{m_{\tilde{a}}}.
\eea
Clearly, this decay width dominates over the one in Eq.~(\ref{decay2bodyaxino}), and therefore the axino lifetime can be approximated as
\begin{equation}
\tau_{\tilde{a}}\simeq \Gamma^{-1} \left( \tilde{a} \rightarrow \psi_{3/2} \, a \right) \simeq 1.18 \times 10^{13} s \, \left( \frac{m_{3/2}}{0.1 \text{ GeV}} \right)^2 \, \left( \frac{1 \text{ GeV}}{m_{\tilde{a}}} \right)^5,
\end{equation}
where to write the second equality we have neglected the contribution of $r_{3/2}$ in Eq.~(\ref{decayto32a}) which is valid when $m_{3/2}\ll m_{\tilde{a}}$.

\subsection{Relic density for multicomponent dark matter}
\label{sec:ddm}

Unlike the gravitino whose lifetime is much longer than the age of the Universe, 
the axino has a smaller lifetime as shown in the previous subsection, and thus one has to consider that its density changes in time with the result
\bea
\Omega_{\tilde{a}}h^2 = \Omega_{\tilde{a}}^{\text{TP}}h^2 e^{-(t_{today}-t_0)/ \tau_{\tilde{a}}},
\label{NLSPrelicgeneral}
\eea
where $t_0$ is the time when the axinos are thermally produced, and $\Omega_{\tilde{a}}^{\text{TP}}h^2$ corresponds to the would-be axino NLSP relic density if it were stable and would not undergo through
the decay process. This relic density depends heavily on the axion model considered. Here we will work in the framework of the KSVZ model~\footnote{The case for axino LSP as the only DM component was discussed in Ref.~\cite{GomezVargas:2019vci} for the KSVZ and DFSZ~\cite{Dine:1981,Zhitnitsky:1980} axion models. In the DFSZ framework, the axino production is dominated by the axino-Higgs-higgsino and/or the axino-quark-squark interactions, and 
its relic density turns out to be in a good approximation independent of the reheating temperature $T_R$, unlike the KSVZ model (see Eq.(\ref{relicaxinos})). However, the axino decay into photon plus neutrino and gravitino plus axion is model independent. In Ref.~\cite{GomezVargas:2019vci}, one can see that the DFSZ allowed region represents a subset of the KSVZ region, as expected, since the former model has one less degree of freedom, $T_R$, in order to obtain the correct relic density. Therefore, this work will be focused on the KSVZ model to explore axino DM with a broad approach.}~\cite{Kim:1979,Shiftman:1980}, where the axino production is dominated by the scatterings of gluons and gluinos, thus its relic density from thermal production is~\cite{Brandenburg:2004,Strumia:2010}
\begin{equation}
\Omega_{\tilde{a}}^{\text{TP}}h^2\simeq 0.3\ (g_3(T_R))^4\ \left(\frac{F(g_3(T_R))}{23}\right) \left( \frac{m_{\tilde{a}}}{1 \text{ GeV}} \right) \left(\frac{T_R}{10^4 \text{ GeV}}\right) \left(\frac{10^{12}\text{ GeV}}{f_a}\right)^2,
\label{relicaxinos}
\end{equation}
where $T_R$ is the reheating temperature after inflation, $g_3$ is the running $SU(3)$ coupling, the rate function $F(g_3(T_R))$ describes the axino production rate with $F\simeq 24-21.5$ for $T_R\simeq 10^4-10^6$ GeV~\cite{Strumia:2010}. For our numerical computation we will use 
$F\simeq 23$. Other values will not change significantly the final results.
Assuming the conservative limit $T_R \gtrsim 10^4$ GeV, 
an upper bound for $m_{\tilde a}$ is obtained for each value of $f_a$ from the measured value of the relic density by the
Planck Collaboration~\cite{Aghanim:2018eyx}
$\Omega_{cdm}^{\text{Planck}}h^2\simeq 0.12$.
For example, 
one obtains $m_{\tilde{a}} \lesssim 50, 0.5, 0.005$ GeV for $f_a=10^{13}, 10^{12}, 10^{11}$ GeV, respectively.



To compute now {the} gravitino relic density, we need to consider thermal and non-thermal production mechanisms. The latter, in our multicomponent scenario, is related to the decay of the axino NLSP. 
Taking all the above into account, the density for gravitino LSP is given by:
\bea
\Omega_{3/2}h^2 = \Omega_{3/2}^{\text{TP}}h^2  + \Omega_{3/2}^{\text{NTP}}h^2,
\label{LSPrelicgeneral}
\eea
where
\bea
\Omega_{3/2}^{\text{NTP}}h^2 = r_{3/2} \, \Omega_{\tilde{a}}^{\text{TP}}h^2 \, (1-e^{-(t_{today}-t_0)/ \tau_{\tilde{a}}}),
\label{LSPrelicNTP}
\eea
and~\cite{Bolz:2000fu,Rychkov:2007uq}:
\begin{equation}
\Omega_{3/2}^{\text{TP}}h^2\simeq 0.02\left(\frac{T_R}{10^5 \text{ GeV}}\right)\left(\frac{1 \text{ GeV}}{m_{3/2}}\right)\left(\frac{M_3(T_R)}{3\text{ TeV}}\right)^2\left(\frac{\gamma / (T_R^6/M_P^2)}{0.4}\right).
\label{relicgravitinos}
\end{equation}
Here
$M_3(T_R)$ is the running gluino mass, and the last factor parametrizes the effective production rate ranging $\gamma(T_R) / (T_R^6/M_P^2)\simeq 0.4-0.35$ for $T_R\simeq 10^4-10^6$~GeV~\cite{Rychkov:2007uq}.
For our numerical computation we will use $M_3(T_R) \simeq 3$ TeV and $\gamma(T_R)/(T_R^6/M_P^2) \simeq 0.4$. Other values will not modify significantly our results.
Assuming as before $T_R \gtrsim 10^4$ GeV, a lower limit for the gravitino mass 
is obtained, $m_{3/2} \gtrsim 0.017 \text{ GeV}$.
Since the gravitino is the LSP, note that this limit is not compatible with the bound for the axino NLSP mass
$m_{\tilde{a}} \lesssim 0.005$ GeV corresponding to $f_a=10^{11}$ GeV, thus we will work with
$f_a\geq 10^{12}$ GeV.


Obviously, if $\tau_{\tilde{a}}\ll t_{today}$, we get the usual relations~\cite{Covi:1999ty,Choi:2011yf,Roszkowski:2014}
\bea
\Omega_{\tilde{a}}h^2 &\simeq& 0,\\
\Omega_{3/2}h^2 &\simeq& \Omega_{3/2}^{\text{TP}}h^2 + r_{3/2} \, \Omega_{\tilde{a}}^{\text{TP}}h^2.
\label{relicgrav2}
\eea

To continue we must address the axion production. Topological defects known as cosmic strings and domain walls can appear in models with axions arising from the breaking of a U(1) symmetry~\cite{Kibble:1976sj,Sikivie:2006ni}. The decay of these defects could be an important source of axions. Nevertheless, as we consider $f_a > T_R$, the PQ symmetry 
is broken before the end of inflation and therefore we assume that the contributions from topological defects are diluted away during inflation. Then, the relevant contributions come from the misalignment mechanism and the axino NLSP decay. For the {former production}, 
the axion cold DM relic density can be accounted by~\cite{Turner:1985si,Sikivie:2006ni,Visinelli:2009zm}
\begin{equation}
\Omega_{a}h^2\simeq 0.18 \theta^2_i \left(\frac{f_a}{10^{12}\text{ GeV}}\right)^{1.19},
\label{relicaxions}
\end{equation}
where $\theta_i$ is the initial misalignment angle. Since we are interested in studying scenarios with axino-gravitino as the only two components of the DM, we can set the axion primordial relic negligible choosing an appropriated value for $\theta_i$ if needed, 
i.e. when $f_a\gsim 10^{12}$~GeV. 
Nevertheless, it would be convenient to work with the upper bound $f_a\leq 10^{13}$ GeV to avoid too much tuning.

Taking all the above discussions into account, throughout this work we will adopt the following range for the PQ scale:
\begin{equation}
10^{12} \leq f_a \leq 10^{13}\ \text{GeV}.
\label{pqscale}
\end{equation}

On the other hand, 
the axions produced by the axino NLSP decay will constitute `dark radiation', i.e., ultrarelativistic and invisible species with respect to the cold DM measured by Planck. The amount of dark radiation is under stringent constraints~\cite{Poulin:2016,Berezhiani:2015,Chudaykin:2016,Chudaykin:2017,Bringmann:2018}, and as a consequence it gives a small contribution to the total DM density.
A quantity that will be useful along this work {to compare with the experimental bounds,} is the fraction of the axino NLSP that decays into dark radiation. For that we can define
\bea
f_{ddm}^{\text{DR}} = f_{\tilde{a}} \left( 1 - r_{3/2} \right),
\label{ddmfraction}
\eea
with
\bea
f_{\tilde{a}}=\frac{\Omega^{\text{TP}}_{\tilde{a}}}{\Omega_{cdm}^{\text{Planck}}}
\label{fractionaxino}
\eea
as the axino NLSP fraction. The subscript $ddm$ denotes decaying dark matter, and $\text{DR}$ stands for dark radiation. Note that $f_{ddm}^{\text{DR}}$ represents the total cold DM that could be transferred into radiation, i.e. the amount of axino that would not produce non-thermal gravitinos, assuming that all the axinos have already decayed, $f_{ddm}^{\text{DR}} = \frac{\Omega^{\text{TP}}_{\tilde{a}}h^2 - \Omega^{\text{NTP}}_{3/2}h^2}{\Omega_{cdm}^{\text{Planck}}}$, with $\Omega^{\text{NTP}}_{3/2}h^2 = r_{3/2} \Omega^{\text{TP}}_{\tilde{a}}h^2$  as in 
Eq.~(\ref{relicgrav2}).

It is worth mentioning the following:
\begin{itemize}
\item Planck obtains $\Omega_{cdm}^{\text{Planck}}h^2\simeq 0.12$ today from measurements at recombination time using the standard $\Lambda$CDM model. We are working with decaying DM, so the cold DM density has a time dependence due to the fact that some of the axino NLSP energy density is `lost' as dark radiation. Nevertheless, the latter quantity has to be small, as discussed above.
\item Decaying DM and its fraction to dark radiation, $f_{ddm}^{\text{DR}}$, refers to the contribution of the mentioned decay of axino NLSP into gravitino LSP plus axion, not to be confused with the decays of axino NLSP and gravitino LSP into photon plus neutrino.
\end{itemize}

Let us finally point out that due to the axion-photon mixing, the axions emitted from the axino decay can be converted into photons in the presence of a magnetic field, potentially producing a signal. {In Ref.~\cite{Bae:2019}, the authors studied possible signatures, for two decaying DM scenarios with axion and axion-like particles (ALPs) as dark radiation. They compared the expected photon flux considering the Galactic magnetic field with the observed diffuse $\gamma$-ray flux by \Fermi-LAT. Although, the axion (and the ALP) to photon conversion probability depends on the DM profile and the Galactic magnetic field profile, for a QCD axion (as in our case), they found that the conversion probability is too small to be observed.}


\subsection{$\gamma$-ray flux from gravitino and axino decays}
\label{sec: gammaflux}

 The differential flux of $\gamma$ rays from DM decay in the Galactic halo is calculated by integrating its distribution around us along the line of sight: 
 \begin{equation}
  \frac{d\Phi_{\gamma}^{\text{halo}}}{dEd\Omega}=\frac{1}{4\,\pi\,\tau_{\textit{DM}}\,m_{\textit{DM}}}\,\frac{dN^{\text{total}}_{\gamma}}{dE} \, \frac{1}{\Delta\Omega}\,
\int_{\Delta\Omega}\!\!\cos
  b\,db\,d\ell\int_0^{\infty}\!\! ds\,\rho_{\text{halo}}(r(s,\,b,\,\ell))\ ,
  \label{eq:decayFlux}
 \end{equation}
 where $\tau_{\textit{DM}}$, $m_{\textit{DM}}$ are the lifetime and mass of the DM particle respectively, $\frac{dN^{\text{total}}_{\gamma}}{dE}$ is the total number of photons produced in a DM decay, $\Delta \Omega$ is the region of interest (ROI), i.e. the region of the sky we are studying, $b$ and $\ell$ denote the Galactic latitude and longitude, respectively, and $s$ the distance from the Solar System. The radius $r$ in the DM halo density profile of the Milky Way, $\rho_{\text{halo}}$, is expressed in terms of these Galactic coordinates.

The constraints to the $\gamma$-ray emission from DM decay are usually presented as lower limits to the particle lifetime, considering that the DM is composed by only one particle species. If gravitino and axino coexist, being one the LSP an the other the NLSP, respectively, both candidates can be sources of $\gamma$-ray radiation. 
As discussed recently in Ref.~\cite{GomezVargas:2019vci}, in a multicomponent scenario it is useful to assume an effective lifetime to normalize the signal considering that a specific source is a fraction of 
$\Omega_{cdm}^{\text{Planck}}$. Assuming that the distribution of each species is homogeneous along the DM distribution, for the $i$-th DM component we can define
\begin{equation}
\tau_{\text{DM}_i\text{-eff}}= f_{\text{DM}_i}^{-1} \, \tau_{\text{DM}_i}, \hspace{1cm} \text{with} \hspace{1cm} f_{\text{DM}_i} = \frac{\Omega_{\text{DM}_i}}{\Omega_{cdm}^{\text{Planck}}},
 \label{newtime1}
\end{equation}
where $f_{\text{DM}_i}$ is the $i$-th DM component fraction, $\tau_{\text{DM}_i}$ is the inverse of the decay width to photons, and the effective lifetime $\tau_{\text{DM}_i\text{-eff}}$ can be tested against the lower limit reported by the experimental collaborations.

However, we cannot apply straightforwardly the above formulas to our multicomponent DDM scenario made of gravitino LSP ($\text{DM}_1$) and axino NLSP ($\text{DM}_2$). The reason is that their fractions change in time due to axino decay into gravitino, so taking into account Eqs.~(\ref{NLSPrelicgeneral}) and~(\ref{LSPrelicgeneral}), we must do the following replacements in Eq.~(\ref{newtime1}) for axino and gravitino respectively:
\bea
 f_{\text{DM}_2} & \rightarrow & f_{\tilde{a}} \, e^{-(t_{today}-t_0)/\tau_{\tilde{a}}},
\label{nueva}
\\
 f_{\text{DM}_1} & \rightarrow & f_{3/2} + r_{3/2} \, f_{\tilde{a}} \, \left(1 - e^{-(t_{today} - t_0)/\tau_{\tilde{a}}} \right),
\label{fractionstime}
\eea
with $f_{\tilde{a}}$ given by Eq.~(\ref{fractionaxino}) and
\bea
f_{3/2}=\frac{\Omega_{3/2}^{\text{TP}}}{\Omega_{cdm}^{\text{Planck}}}.
\eea
As expected, if axino NLSP decay into gravitino LSP plus axion is not allowed, one gets the same result as in Eq.~(\ref{newtime1}).

Finally, in a same fashion stated before, it is easier for the analysis to consider an effective lifetime in our multicomponent DDM scenario. Thus Eq.~(\ref{newtime1}) becomes
\bea
 \tau_{\tilde{a}\text{-eff}} & = & \left( f_{\tilde{a}} \, e^{-(t_{today}-t_0)/\tau_{\tilde{a}}} \right)^{-1} \, \Gamma^{-1}\left( \tilde{a} \rightarrow \gamma \nu_i \right) ,
\label{newtime2}
\\
 \tau_{3/2\text{-eff}} & = & \left[ f_{3/2} + r_{3/2} \, f_{\tilde{a}} \, \left(1 - e^{-(t_{today} - t_0)/\tau_{\tilde{a}}} \right) \right]^{-1} \, \Gamma^{-1}\left( \psi_{3/2} \rightarrow \gamma \nu_i \right).
\label{efflifetimes}
\eea

It is now straightforward to apply the analyses of these Subsections to study the current constraints on the parameter space of our scenario, as well as the prospects for its detection. For simplicity, in what follows we will use $t_0 = 0$ for the computation.

\section{Results}
\label{sec: results}

\subsection{Constraints from cosmological observations}
\label{sec: cosmoobstconstraints}



\begin{figure}[t!]
\begin{center}
 \begin{tabular}{cc}
 \hspace*{-4mm}
 \epsfig{file=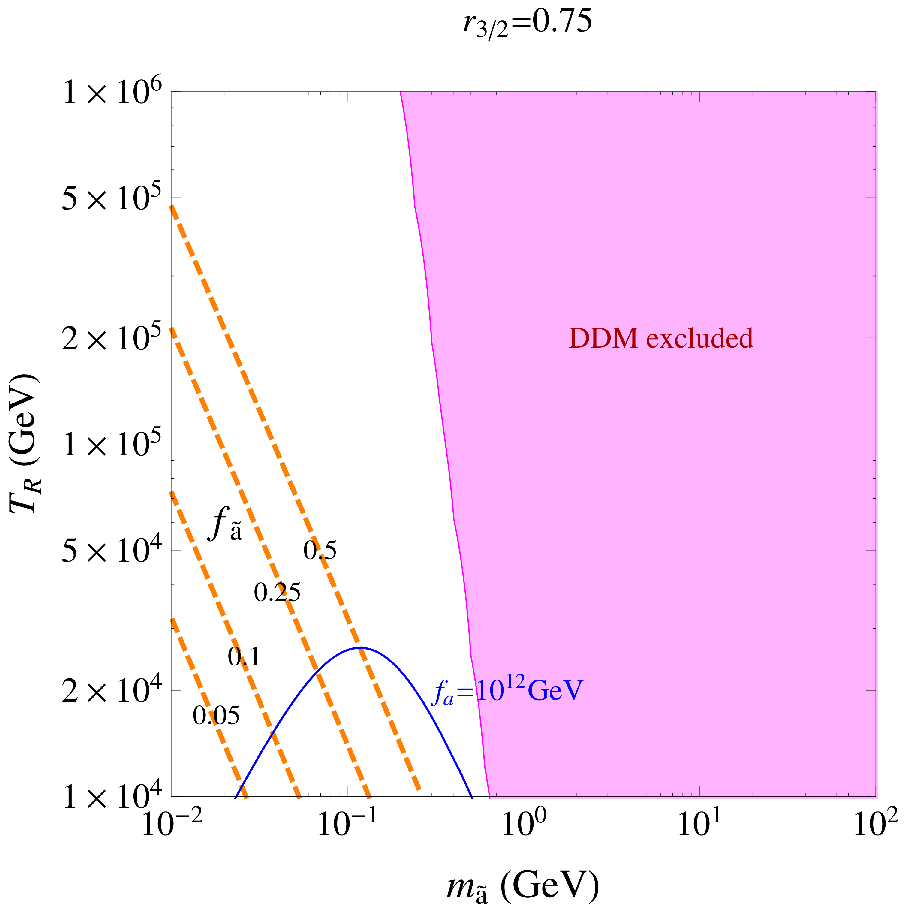,height=7cm} 
       \epsfig{file=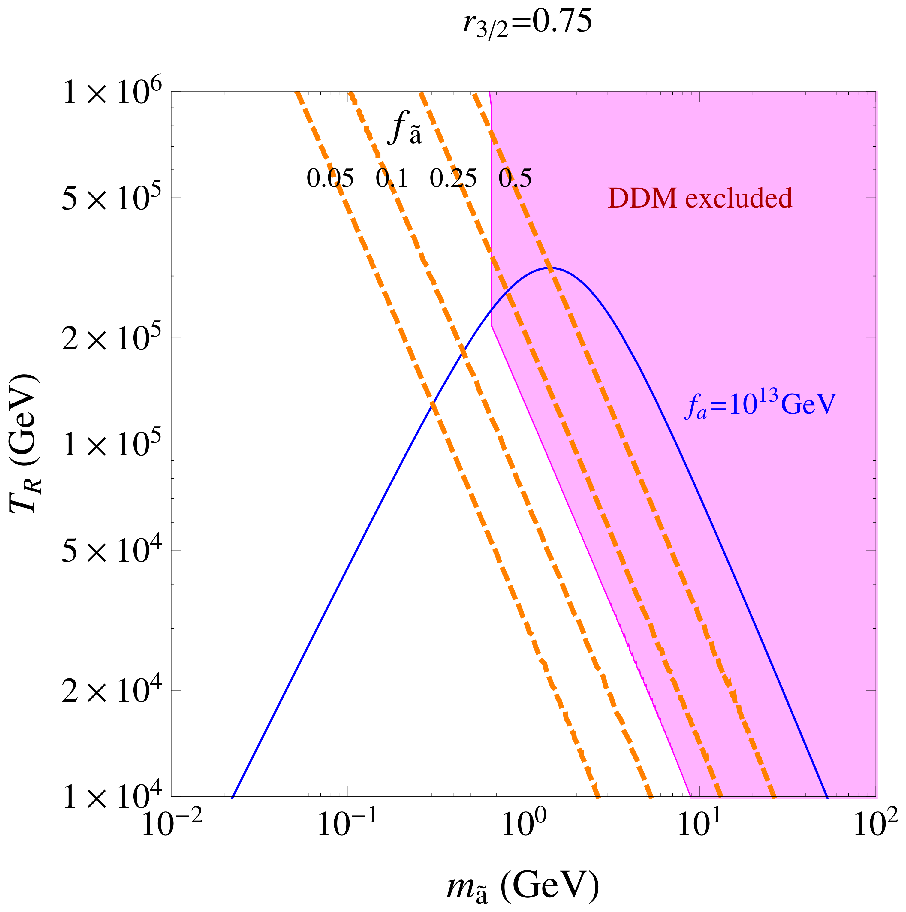,height=7cm}   
       \vspace*{-0.8cm}       
   \\ & 
    \end{tabular}
    \captions{Constraints on the reheating temperature versus {the} axino NLSP mass for the multicomponent DDM scenario with gravitino LSP, 
and {a} mass relation $r_{3/2}=0.75$. Blue lines corresponds to points with $\Omega_{3/2}h^2+\Omega_{\tilde{a}}h^2$ equal to $\Omega_{cdm}^{\text{Planck}}h^2$ at recombination era in agreement with Planck observations, for two values of the PQ scale $f_a=10^{12}$ GeV (left panel) and $10^{13}$ GeV (right panel). The regions above the blue lines are excluded by overproduction of cold DM. The magenta region is excluded by cosmological observations for DDM models~\cite{Poulin:2016,Berezhiani:2015,Chudaykin:2016,Chudaykin:2017,Bringmann:2018}, considering bounds on $f_{ddm}^{\text{DM}}$. Orange dashed lines correspond to the axino NLSP fractions $f_{\tilde{a}}=0.5, 0.25, 0.1, 0.05$. 
The upper bounds 
$m_{\tilde{a}} \lesssim 0.5, 50$ GeV in left and right panels, respectively,
are obtained from Eq.~(\ref{relicaxinos}) assuming the conservative limit $T_R \gtrsim 10^{4}$ GeV.
}
    \label{figgravitinoLSP1}
\end{center}
\end{figure}

To analyze the regions of the parameter space that can satisfy the current experimental constraints on DDM models, similar as in Ref.~\cite{GomezVargas:2019vci} we show $T_R$ versus $m_{\tilde a}$ for a fixed $r_{3/2}=0.75$ in Fig.~\ref{figgravitinoLSP1}. 
The left panel corresponds to the PQ scale 
$f_a=10^{12}$ GeV, whereas the right panel to $f_a=10^{13}$ GeV. 
We will also remark some differences with the results of Ref.~\cite{GomezVargas:2019vci}, where the opposite situation, axino LSP and gravitino NLSP was analyzed.


The blue lines show points of the parameter space with $\Omega_{3/2}h^2+\Omega_{\tilde{a}}h^2$ fulfilling Planck observations at recombination era. 
The regions above the blue lines are excluded by overproduction of cold DM. The region below them could be allowed if we assume a third DM contribution, but for simplicity
we will focus on values of the parameters fulfilling the blue contours.
On the other hand,
the orange dashed lines correspond to different values of the axino NLSP fraction 
$f_{\tilde{a}}$.

The magenta regions in both panels are excluded by cosmological observations for DDM models~\cite{Poulin:2016,Berezhiani:2015,Chudaykin:2016,Chudaykin:2017,Bringmann:2018}, taking into account the stringent constraints on the fraction of axino NLSP relic density that decays to dark radiation, $f_{ddm}^{\text{DM}}$. The bounds considered are taken from Ref.~\cite{Poulin:2016} at 95\% CL,
\bea
f_{ddm}^{\text{DM}} \, \Gamma(\tilde{a} \rightarrow \psi_{3/2} \, a) < 15.9 
\times 10^{-3}\ \text{Gyr}^{-1} \hspace{1cm} &\text{if}& \hspace{0.5cm} \tau_{\tilde{a}} > t_{today},  \\
f_{ddm}^{\text{DM}} < 0.042 \hspace{1cm} &\text{if}& \hspace{0.5cm} t_{today} > \tau_{\tilde{a}} > t_{rec},
\eea
where $t_{rec}$ denotes the recombination time. Constraints for DDM with shorter lifetimes can be found in Fig.~5 of that work. We derive the excluded regions for our model using a grid of values of $m_{\tilde{a}}$ and $T_R$. If $t_{today} > \tau_{\tilde{a}} > t_{rec}$, the $f_{ddm}^{\text{DM}}$ upper limit translates into an upper limit on $f_{\tilde{a}}$, for a fixed $r_{3/2}$, according to Eq.~(\ref{ddmfraction}). This can be seen in the right panel of Fig.~\ref{figgravitinoLSP1} for $m_{\tilde{a}}\gtrsim 0.6$ GeV.
For axino decaying after the present era, the experimental bound cannot be expressed as a constant upper limit on $f_{\tilde{a}}$, as it involves the product of $f_{ddm}^{\text{DM}}$ and $\Gamma(\tilde{a} \rightarrow \psi_{3/2} \, a)$.
This can be seen in the left panel of Fig.~\ref{figgravitinoLSP1} for $m_{\tilde{a}} < 0.6$ GeV,  but note that the blue line is not affected by this bound.
%
%


Note that unlike Ref.~\cite{GomezVargas:2019vci} for the axino LSP case, where we can show for a given $r_{\tilde a}$ several blue lines, corresponding to different values of $f_a$, with the corresponding DDM excluded region, here we cannot do the same for a given $r_{3/2}$. The reason being that 
the axino thermal relic density, Eq.~(\ref{relicaxinos}), depends on the PQ scale, and therefore the DDM constraints change when we change the axino relic density fraction $f_{\tilde{a}}$.

Another important difference is that here for higher LSP masses, the NLSP relic density increases.
However, in the case discussed in Ref.~\cite{GomezVargas:2019vci}, for higher LSP masses the NLSP relic density decreases. 
This modifies the shape of the DDM exclusion region, because the NLSP decay is the source of the ultrarelativistic particles. 
The left panel of Fig.~\ref{figgravitinoLSP1} depicts a long-lived axino (the axino decay into gravitino plus axion takes place after the present era), whereas the right panel an intermediate-lived axino (the decay takes place between recombination and the present era). 

\begin{figure}[t!]
\begin{center}
 \begin{tabular}{cc}
 \hspace*{-4mm}
 \epsfig{file=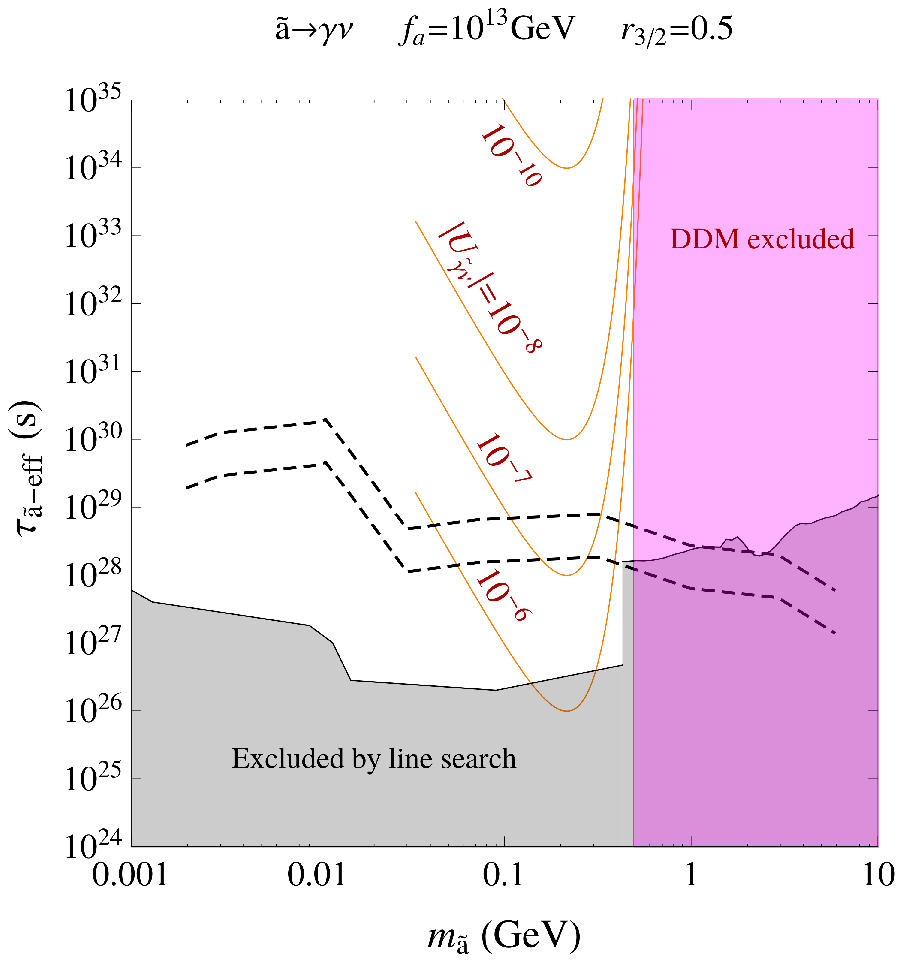,height=7cm}
       \epsfig{file=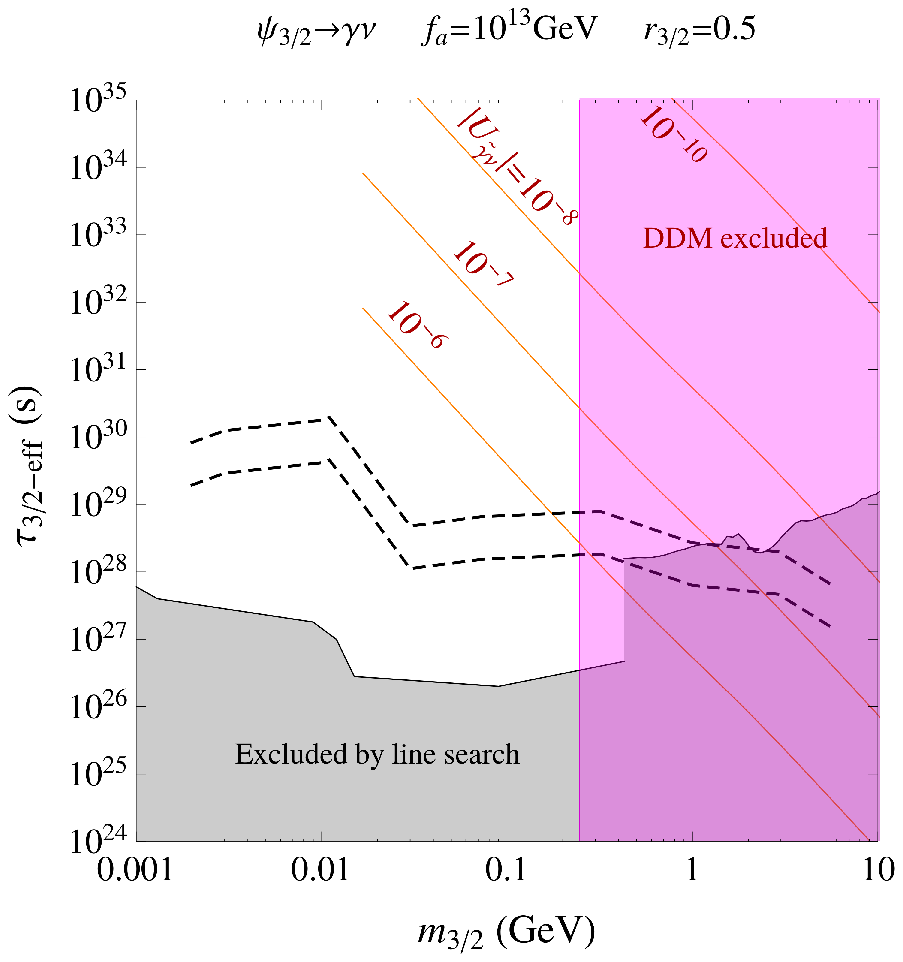,height=7cm}  
       \vspace*{-0.8cm}      
   \\ &
    \end{tabular}
    \captions{
    Constraints on {the} effective lifetime versus {the} axino NLSP mass (left panel) 
and {the} gravitino LSP mass (right panel). The $\gamma$-ray signals for {the} axino and gravitino decays are analyzed separately in left and right panels, respectively. {The grey region below the black solid line is excluded by line searches in the Galactic halo by 
COMPTEL~\cite{Essig:2013goa} (leftmost limit) and \Fermi LAT~\cite{Ackermann:2015lka} (rightmost limit) at 95\% CL.} 
The region below the upper (lower) black dashed line could be probed by e-ASTROGAM~\cite{eAstrogamAngelis:2017} with observations of {a ROI of  
10$^\text{o} \times$10$^\text{o}$
around the Galactic center}, assuming Einasto B (Burkert) DM profile. The orange solid lines correspond to the predictions of the $\mu \nu$SSM for several representative values of $|U_{\tilde{\gamma} \nu}|$.
The lower bound $m_{3/2} \gtrsim 0.017$ GeV is obtained from 
Eq.~(\ref{relicgravitinos}) assuming the conservative limit $T_R \gtrsim 10^4$ GeV. The magenta region is excluded by cosmological observations for DDM models~\cite{Poulin:2016,Berezhiani:2015,Chudaykin:2016,Chudaykin:2017,Bringmann:2018}, considering the bound on $f_{ddm}^{\text{DR}}$.
}
    \label{figgravitinoLSP2}
\end{center}
\end{figure}

\subsection{Constraints~from~$\gamma$-ray~observations~and~prospects~for~detection}
\label{detection}

To analyze the effect on $\gamma$-ray searches of axino NLSP decaying into gravitino LSP, in Fig.~\ref{figgravitinoLSP2} we show the effective lifetime versus the DM candidate mass for one example of the parameter region, $f_a=10^{13}$ GeV and $r_{3/2}=0.5$, where a double-line signal could be detected.
The left (right) panel shows the limits on the parameter space considering the line produced by axino NLSP (gravitino LSP) decaying into $\gamma \nu$. 
Thus the two panels correspond to the same DDM scenario, and the constraints obtained from both of them have to be taken into account for each point of the parameter space. 

The magenta regions are excluded by cosmological observations concerning dark radiation.
The grey regions below the black solid lines are excluded by line searches 
by COMPTEL~\cite{Essig:2013goa} and \Fermi LAT~\cite{Ackermann:2015lka} {at 95\% CL}. The black dashed lines correspond to the projected e-ASTROGAM sensitivity~\cite{eAstrogamAngelis:2017}. {To estimate the bounds}, we have considered the following DM profiles for the observations of a ROI of 10$^\text{o} \times$10$^\text{o}$ around the Galactic center: NFW~\cite{Navarro:1995iw}, Moore~\cite{Diemand:2004wh}, Einasto~\cite{Graham:2005xx,Navarro:2008kc}, Einasto B~\cite{Tissera:2010} and Burkert~\cite{Burkert:1995yz}.
In particular, Einasto B (Burkert) is the most (least) stringent and corresponds in 
the figure 
to the upper (lower) dashed line. 
{More details about the computation of the expected sensitivity of e-ASTROGAM
to DM decay to photons, can be found in Appendix~\ref{lineASTROGAM}.}

Using the results from previous subsections, we also show
in Fig.~\ref{figgravitinoLSP2} with orange solid lines the values of the parameters predicted by the $\mn$ 
for several representative values of $|U_{\tilde{\gamma} \nu}|$.
{Let us remark that the points between the orange curves reproduce the correct relic density as a combination of axinos and gravitinos, for different values of $T_R$, as discussed in Fig.~\ref{figgravitinoLSP1}.}
In the left panel, which
represents the limits considering the line produced only by {the} axino NLSP decaying into $\gamma \nu$,  
we can see the effect of the reduction of the axino relic density due to its decay 
into {the} gravitino LSP for $m_{\tilde{a}} \gtrsim 0.2$ GeV,
as can be deduced from Eq.~(\ref{newtime2}).
The right panel considers the same parameter space, but analyzing the line produced by the gravitino LSP decaying into $\gamma \nu$. This case has a larger effective lifetime with respect to the case with only 
gravitino DM~\cite{Choi:2009ng,Albert:2014hwa,GomezVargas:2017}, due to the non-thermal contribution discussed.

As we can see in the figure, significant regions below the dashed lines could be probed.
This is specially true thanks to the line signal coming from {the} axino NLSP for 
$0.04\lsim m_{\tilde{a}} \lsim 0.5$ GeV and  $10^{-8} < |U_{\tilde{\gamma} \nu}| \leq 10^{-6}$ (see the left panel).
Moreover, for this example there is also a narrow region in the right panel corresponding to a detectable line signal coming from {the} gravitino LSP, for
$0.15 \lsim m_{3/2} \lsim 0.25$ GeV and $|U_{\tilde{\gamma} \nu}|\approx 10^{-6}$. 
Given the parameter used in the figure, $r_{3/2}=0.5$, this gravitino mass range
corresponds to axino masses
$0.3 \lsim m_{\tilde{a}} \lsim 0.5$ GeV, which are embedded in the range producing a line signal from axino, thus we expect a detectable double line as 
an overwhelming smoking gun of this parameter region.

\begin{figure}[t!]
 \begin{center}
  \begin{tabular}{cc}
 \hspace*{-0mm}
 \includegraphics[height=6cm]{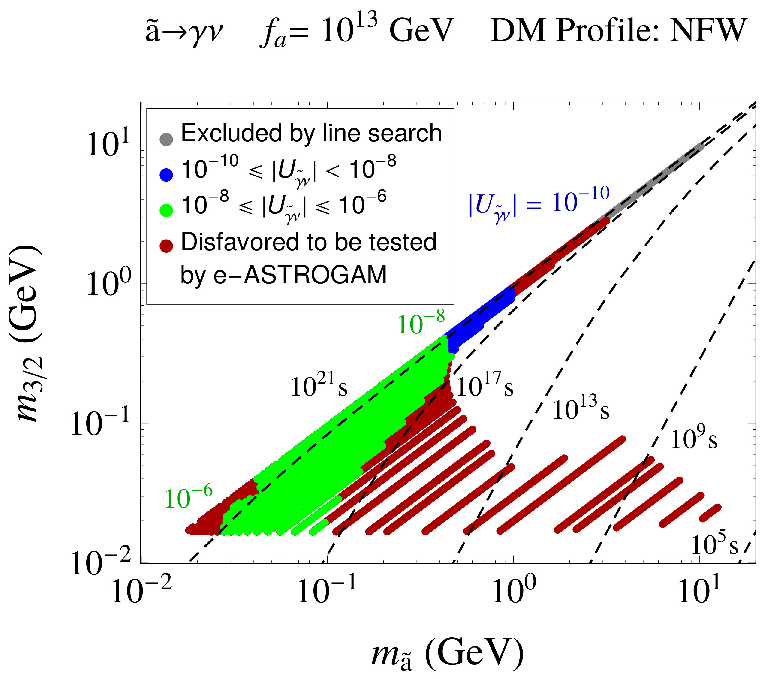} \hspace*{-0.5cm} \includegraphics[height=6cm]{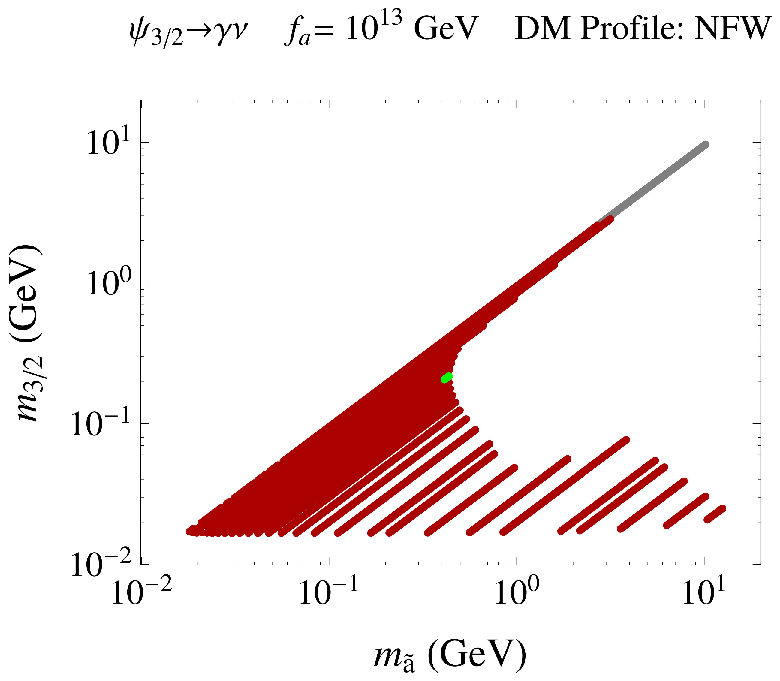}\\     
	\hspace*{-0mm} \includegraphics[height=6cm]{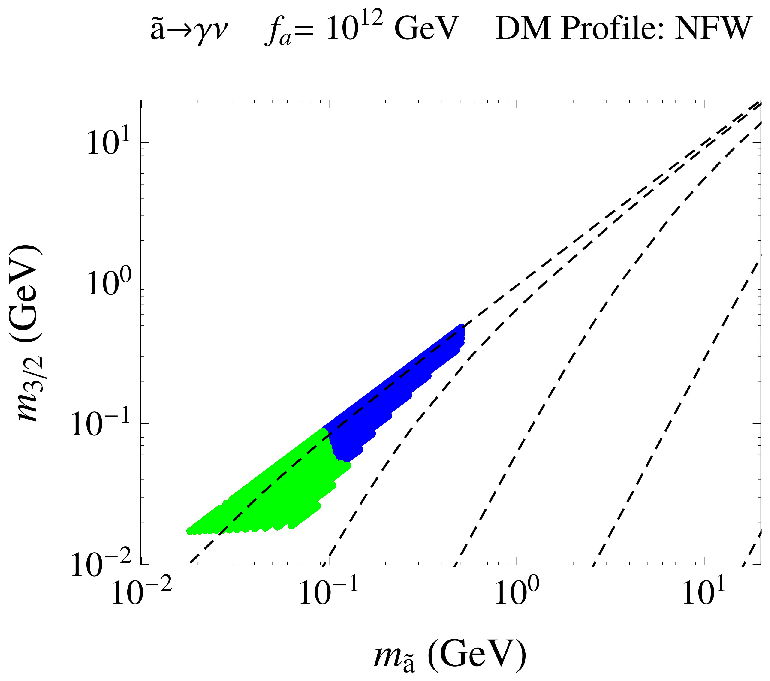} \hspace*{-0.5cm} \includegraphics[height=6cm]{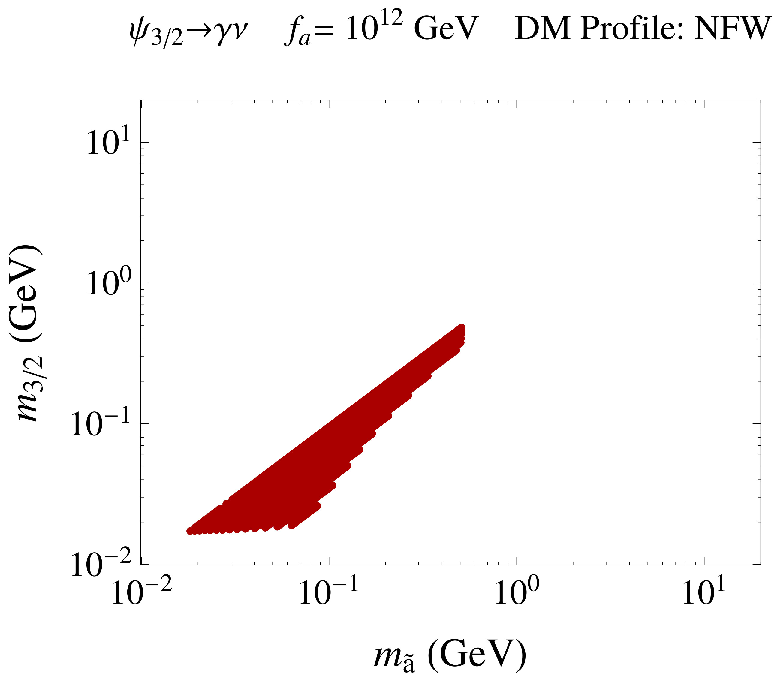}\\ 
  \end{tabular}
  \captions{Constraints on {the} gravitino LSP mass versus {the} axino NLSP mass, with
{the lower bound $m_{3/2} \gtrsim 0.017$ GeV obtained from 
Eq.~(\ref{relicgravitinos}) assuming the conservative limit $T_R \gtrsim 10^4$ GeV.}
The $\gamma$-ray signals from axino and gravitino decays are analyzed separately in left and right panels, respectively, assuming a NFW profile. The grey region corresponds to points excluded by line searches in the Galactic halo by COMPTEL~\cite{Essig:2013goa} and \Fermi LAT~\cite{Ackermann:2015lka} {at 95\% CL}. Blue and green regions correspond to points that could be probed by e-ASTROGAM for the representative ranges $10^{-10} \leq |U_{\tilde{\gamma} \nu}| < 10^{-8}$ and $10^{-8} \leq |U_{\tilde{\gamma} \nu}| \leq 10^{-6}$, respectively, in the $\mu \nu$SSM. In the top left panel, the values in the border between regions are labeled, and for the rest of the panels the labeling is the same. If the same point can be probed in both panels, a double-line signal could be measured. The red region corresponds to points disfavored to be tested by e-ASTROGAM. In the left panels, 
the black {dashed} lines show different values of $\tau_{\tilde{a}} \simeq \Gamma^{-1}(\tilde{a}\rightarrow \psi_{3/2} \, a)$ between $10^5$ and $10^{21}$ s. All the points shown satisfy $\Omega_{3/2}h^2+\Omega_{\tilde{a}}h^2$ equal to $\Omega_{cdm}^{\text{Planck}}h^2$ at recombination era in agreement with Planck observations, as well as DDM constraints for $f_{ddm}^{\text{DR}}$.
}
    \label{figconstrains2}
\end{center}
\end{figure}

To carry out now the complete analysis of the allowed parameter space, for $m_{3/2} \gtrsim 0.017$ GeV {we have performed a scan over the ranges: $1.05\ m_{3/2} \lsim m_{\tilde a} \lsim 50, 0.5$ GeV for $f_a=10^{13}, 10^{12}$ GeV, respectively, taking into account that axino is always heavier than gravitino and the bounds on their masses discussed below 
Eqs.~(\ref{relicaxinos}) and~(\ref{relicgravitinos}).}
The result is shown in Fig.~\ref{figconstrains2}, where the $\gamma$-ray signals from axino and gravitino decays are analyzed separately in the left and right panels, respectively. Green and blue regions correspond to points that could be probed with the projected sensitivity of e-ASTROGAM
assuming a NFW profile, 
for different values of the photino-neutrino mixing parameter $|U_{\tilde{\gamma} \nu}|$. In particular, the green points corresponds to the most natural range of $|U_{\tilde{\gamma} \nu}|$ as discussed in 
Eq.~(\ref{relaxing}). It is worth
mentioning here that this range includes the typical parameter space that can reproduce the observed neutrino physics in bilinear RPV models, thus the constraints obtained also apply to those models.



%
As we can see in the figure, for values of $r_{3/2}$ close to 1, i.e. the upper border line,
we recover the allowed parameter space obtained in the 
work~\cite{GomezVargas:2019vci} with axino LSP and gravitino NLSP
if we consider now both gravitino LSP and axino NLSP effect.
{For $f_a=10^{13}$~GeV (top panels) and $3 \lsim m_{\tilde{a}} \lsim 10$ GeV, the excluded grey region is determined by the non-observation of a $\gamma$-ray line coming from axino NLSP. The red points for $1 \lsim m_{\tilde{a}} \lsim 3$ GeV (top left panel), denote a region where a signal produced by axino NLSP is expected, but it is only allowed for $|U_{\tilde{\gamma} \nu}|\lsim 10^{-9}$, where the 
$\tau_{\tilde{a}\text{-eff}}$
is close to the current line constraints. However, the latter mass range would not be probed by e-ASTROGAM, as the estimated sensitivity used in this work lies below the lifetime lower limit set by \Fermi LAT.}

Even though the DDM constraints for $f_{ddm}^{\text{DR}}$ become relaxed 
since 
$\tau_{\tilde a}^{-1}\simeq\Gamma(\tilde{a} \rightarrow \psi_{3/2} \, a) \rightarrow 0$ when $r_{3/2} \rightarrow 1$ (see Eq.~(\ref{decayto32a})), the remaining effect concerning the $\gamma$-ray flux dominates: the initial relic density fractions of the LSP and NSLP that do not change in time, as can be seen from Eqs.~(\ref{nueva}) and~(\ref{fractionstime}), cannot be ignored. 
For the entire mass region, the initial axino NLSP relic density is relevant, and its decay to photon-neutrino can give rise to a signal and set constraints on the parameter space that otherwise would not exist considering only the contribution from gravitino LSP.


From Fig.~\ref{figconstrains2}, we can conclude that a significant region of the parameter space of our DDM scenario could be tested by next generation $\gamma$-ray telescopes. {Similar to the case discussed in Ref.~\cite{GomezVargas:2019vci}, where the axino (LSP) is the main source of the relevant photon signal, here also the axino (NLSP) plays the same role as shown in the left panels, instead of the gravitino LSP as one would expect naively. 
Note in this sense that the black {dashed} lines in the left panels show us
that this photon signal lies in the region of the parameter space with 
$\tau_{\tilde a}\simeq\Gamma^{-1} \left( \tilde{a} \rightarrow \psi_{3/2} \: a \right) > t_{today}$, and, on the other hand, the axino and gravitino decay widths to photon-neutrino which are relevant quantities for the amount of photon flux 
(see Eqs.~(\ref{eq:decayFlux}),~(\ref{newtime2}) and~(\ref{efflifetimes})) 
fulfills always $\Gamma(\tilde{a}\rightarrow\gamma\nu_i) > \Gamma(\psi_{3/2}\rightarrow\gamma\nu_i)$ as discussed in Sect.~\ref{sec: axino NLSP}.
In particular, this detectable region is inside the following mass ranges: $20$ MeV $\lsim m_{\tilde a} \lsim 1$ GeV and $17$ MeV $\lsim m_{3/2} \lsim 1$ GeV.

According to this discussion, we also expect a line signal coming from {the} gravitino LSP to be measured in a smaller region. This is actually the green region of the top right panel corresponding $f_a=10^{13}$~GeV. It is inside the ranges
$370\lsim m_{\tilde a} \lsim 500$ MeV and $180\lsim m_{3/2} \lsim 250$ MeV, with $0.45\lsim r_{3/2} \lsim 0.55$.
Since this region is also probed with a line from {the} axino NLSP, and the features from each candidate turn out to be located at different enough energies to do not overlap (notice the values of $r_{3/2}$), we expect that a double-line signal could be measured as an overwhelming smoking gun.
Note that this is the same region already discussed in the example
of Fig.~\ref{figgravitinoLSP2}.}

Although in Fig.~\ref{figconstrains2} we used the projected e-ASTROGAM sensitivity assuming a NFW profile,
we have checked that using a different DM profile, such as Einasto B, the detectable parameter space is not essentially modified.
For the latter profile, we expect a line signal coming from the gravitino LSP inside the mass ranges
$320 \lsim m_{\tilde a}\lsim 500$ MeV and
$150\lsim m_{3/2}\lsim 250$ MeV, for $f_a=10^{13}$~GeV. 

Finally, we would like to point out that we have adopted in this work a conservative approach, using a projected sensitivity for e-ASTROGAM that is far to be optimized for our DDM scenario. The results presented here probe that the experiment is going to be capable to explore the multicomponent DM scenario, including the possibility to detect a double-line signal. Nevertheless, the region where this happens could improve significantly compared
to the one shown here if for instance we used a ROI more efficient for decay 
processes {such as a region of the Galactic halo, as discussed in Appendix~\ref{lineASTROGAM}.}

\section{Conclusions}
\label{sec:conclusions}

In this work, we have analyzed a mixture of gravitino and axino particles as DM, extending a previous work on the subject of multicomponent DM in RPV SUSY~\cite{GomezVargas:2019vci}. Now we have studied the scenario of gravitino LSP with axino NLSP. In this context of DDM, we have found that this combination of particles can reproduce the accumulated DM evidence, avoiding cosmological problems.

In the context of the $\mn$, we have analyzed the possibility of an axino NLSP having a RPC partial decay width into gravitino LSP plus axion, in addition to the RPV partial decay width into photon plus neutrino. The latter decay also occurs for the gravitino LSP, with a lifetime typically much longer than the age of the Universe due to the small values of neutrino Yukawas in the generalized electroweak-scale seesaw of the $\mn$.
If {the} axino and {the} gravitino coexist, both DM particles can be sources of $\gamma$-ray radiation.

The corresponding relic density has been discussed, and assuming a conservative lower bound on the reheating temperature of
$T_R\gsim 10^4$ GeV
an upper bound on the axino mass of $m_{\tilde a} \lsim 50$ GeV was obtained, as well as a lower bound on the gravitino mass of $m_{3/2} \gsim 17$ MeV.
We have also found the regions of the parameter space excluded by cosmological observations, considering the stringent constraints on the fraction of {the} axino NLSP relic density that decays to dark radiation (see Fig.~\ref{figgravitinoLSP1}).

Then, we have studied the $\gamma$-ray flux produced in this DDM scenario 
of the $\mn$, finding that a significant region of the parameter space could be tested by e-ASTROGAM searching in a ROI around the Galactic center. In particular, this region is inside the mass ranges 
$20\ \text{MeV} \lsim m_{\tilde a} \lsim 1$ GeV and $17 \ \text{MeV} \lsim m_{3/2} \lsim 1$ GeV.
This is specially true thanks to the line signal coming from {the} axino 
NLSP decay (see the left panels of Fig.~\ref{figconstrains2}), {instead of a signal produced by the LSP, as one would expect. This has a huge impact, allowing us to probe scenarios with gravitino LSP and $m_{3/2} \lsim 1$ GeV that otherwise would not produce a detectable signal by near future experiments. Naturally, the axino NLSP signature is obtained when}
 $\tau_{\tilde a}\simeq\Gamma^{-1} \left( \tilde{a} \rightarrow \psi_{3/2} \: a \right) > t_{today}$.
 
Additionally, a signal coming from {the} gravitino LSP could be measured in a smaller region of the parameter space for $f_a=10^{13}$~GeV inside the mass ranges
$370\lsim m_{\tilde a} \lsim 500$ MeV and $180\lsim m_{3/2} \lsim 250$ MeV
(see the top right panel of Fig.~\ref{figconstrains2}).
In this case, a double-line signal from {the} axino and gravitino decays could be measured as an overwhelming smoking gun. Let us remark in this sense that this double-line signal is different in energies from the one obtained for the opposite case with gravitino NLSP and axino LSP in 
Ref.~ \cite{GomezVargas:2019vci}.

\section*{Acknowledgments}

The work of GAGV was supported by Programa FONDECYT Postdoctorado under grant 3160153. The work of DL and AP was supported by the Argentinian CONICET, and they also acknowledge the support through PIP 11220170100154CO. 
The work of CM was supported in part by the Spanish Agencia Estatal de Investigaci\'on 
through the grants FPA2015-65929-P (MINECO/FEDER, UE), PGC2018-095161-B-I00 and IFT Centro de Excelencia Severo Ochoa SEV-2016-0597.
We also acknowledge the support of the Spanish Red Consolider MultiDark FPA2017-90566-REDC. 
CM and DL gratefully acknowledge the hospitality of the Institut Pascal during the Paris-Saclay Astroparticle Symposium 2019, supported by P2IO (ANR-11-IDEX-0003-01 and ANR-10-LABX-0038), in whose stay the last stages of this work were carried out.


\appendix

\section{Sensitivity {of e-ASTROGAM to DM decay to photons}}
\label{lineASTROGAM}

The e-ASTROGAM collaboration has presented limits on DM annihilation~\cite{eAstrogamAngelis:2017,Bartels:2017dpb}.
To study the expected sensitivity of e-ASTROGAM to DM decay to photons we must convert the limits on DM annihilation. Here we show how to carry it out.

The differential flux of $\gamma$ rays 
in the Galactic halo can be written as
 \begin{equation}
  \frac{d\Phi_{\gamma}^{\text{halo}}}{dEd\Omega}=\frac{r_{\odot}\, \Gamma}{4\,\pi} \, \left( \frac{\rho_{\odot}}{m_{\textit{DM}}} \right)^{\alpha}\,\frac{dN^{\text{total}}_{\gamma}}{dE} \,\frac{1}{\Delta\Omega} \, \int_{\Delta\Omega}\!\!\cos
  b\,db\,d\ell\int_0^{\infty}\!\! \frac{ds}{r_{\odot}}\,\left(\frac{\rho_{\text{halo}}(r(s,\,b,\,\ell))}{\rho_{\odot}}\right)^{\alpha},
  \label{eq:annFlux}
 \end{equation}
where $\alpha=1$ ($\alpha=2$) corresponds to DM decay (annihilation).
The definition of several of the parameters of Eq.~(\ref{eq:annFlux}) can be found in the
discussion of Eq.~(\ref{eq:decayFlux}), where the differential flux for the case of decay was studied.
From Eq.~(\ref{eq:annFlux}), 
we can identify three important contributions:
\begin{itemize}

\item The DM interaction rate denoted by $\Gamma$. For DM decay, $\Gamma = 1 / \tau_{DM}$. 
For DM annihilation, $\Gamma = a \, \langle \sigma v \rangle$, with $\langle \sigma v \rangle$ the velocity-averaged annihilation cross section, and $a=1/2 \, (1/4)$ if DM is (is not) self conjugated.

\item The photon energy spectrum 
denoted by $\frac{dN^{\text{total}}_{\gamma}}{dE} = N_{\gamma}^{(\alpha)} \, \delta (E- E_{\gamma})$, where $N_{\gamma}^{(\alpha)}$ is the number of photons produced in a single process. For DM decay (annihilation), $E_{\gamma} = m_{DM}/2$ ($E_{\gamma} = m_{DM}$) and we denote $N_{\gamma}^{(1)} = N_{\gamma}^{(dec)}$ ($N_{\gamma}^{(2)} = N_{\gamma}^{(ann)}$).

\item The astrophysical part, i.e. the integral over $\rho_{\text{halo}}$ raised to the power $\alpha$ along the line of sight {divided by $\Delta\Omega$}. 
For DM decay (annihilation), this is called the D (J)-factor.
$\rho_{\odot}$ denotes the DM density at the location of the Sun $r_{\odot}$, and both parameters are included to make the D- and J-factors dimensionless. 
\end{itemize}

The constraints from $\gamma$-ray searches considering DM decay (annihilation) are usually presented as lower limits (upper limits) to the particle lifetime (velocity-averaged annihilation cross section) as a function of the DM mass. Also, the bounds are quoted for observations over a specific ROI, i.e. a particular D-factor (J-factor).

\begin{figure}[t!]
\begin{center}
 \epsfig{file=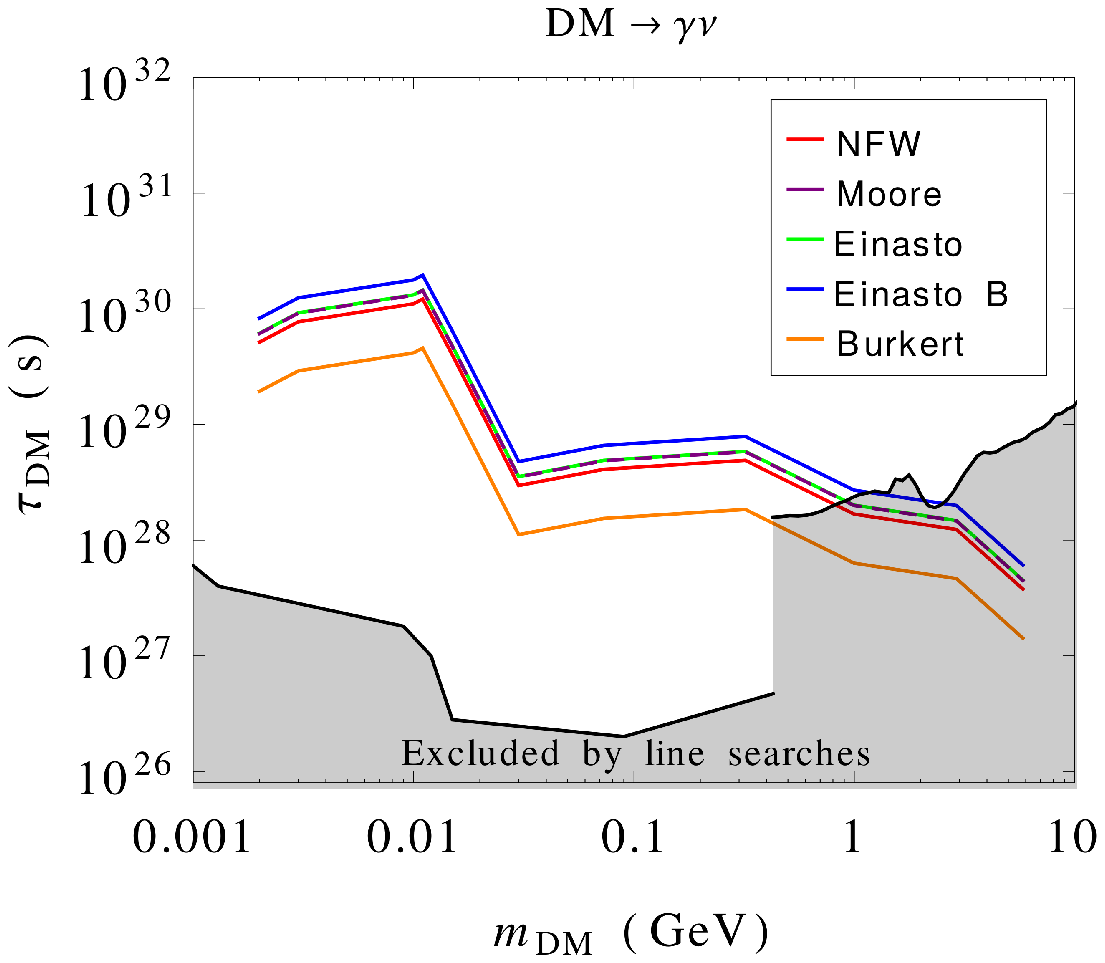,height=7.5
cm}  
    \captions{e-ASTROGAM projected lower limits on DM decay to $\gamma \nu$ for several DM density profiles as indicated in the figure, converting the limits on DM annihilation to $\gamma \gamma$~\cite{eAstrogamAngelis:2017} obtained with a ROI of 10$^\text{o} \times$10$^\text{o}$ around the Galactic center. 
The grey region below the black solid line is excluded by line searches in the Galactic halo by 
COMPTEL~\cite{Essig:2013goa} (leftmost limit) and \Fermi LAT~\cite{Ackermann:2015lka} (rightmost limit) at 95\% CL. 
}
    \label{sASTROGAMsensitivity}
\end{center}
\end{figure}

Note that the signal that can be measured is the same for both processes, decay and annihilation, i.e. a monochromatic $\gamma$-ray line. The only difference 
is that a spectrum with the same energy corresponds to different DM masses. Then, lifetime lower limits and cross-section upper limits can be obtained from the spectral line flux upper limits, as explicitly shown in Ref.~\cite{Ackermann:2012qk}.
Similarly, we can relate the lower limits on $\tau_{DM}$ for DM decay to photons to the upper limits on $\langle \sigma v \rangle$ for DM annihilation to photons (or vice-versa), for the same ROI, as
\bea
\tau_{DM} [m_{DM}] \rightarrow \frac{1}{4 \, a \, \rho_{\odot}} \, \frac{m_{DM}}{\langle \sigma v \rangle [m_{DM}/2]} \, \frac{N_{\gamma}^{(dec)}}{N_{\gamma}^{(ann)}} \, \frac{\text{D-factor}}{\text{J-factor}},
\label{fractionaxino}
\eea
where the notation $\tau_{DM} [m_{DM}]$ and $\langle \sigma v \rangle [m_{DM}/2]$ implies that the bound on the lifetime corresponding to a DM particle with mass $m_{DM}$ is related to the bound on the annihilation cross section corresponding to a DM particle with mass $m_{DM}/2$.

\begin{table}
\begin{center}
\begin{tabular}{ ccccc } 
 \hline
 DM profile & $r_s$ (kpc) & $\rho_s$ (GeV/cm$^3$) & D-factor & J-factor \\ 
  \hline
   \hline
 NFW & 24.42 & 0.184 & 11.8 & 174 \\ 
 Moore & 30.28 & 0.105 & 14.0 & - \\ 
 Einasto & 28.44 & 0.033 & 14.1 & - \\ 
 Einasto B & 35.24 & 0.021 & 19.0 & - \\ 
 Burkert & 12.67 & 0.712 & 4.44 & - \\ 
 \hline
\end{tabular}
 \caption{\label{DMprofilesTable} Values of the parameters $r_s$ and $\rho_s$ for the profiles given in Eq.~(\ref{profiles}), and the corresponding D-factors (and J-factor for the NFW case) for a ROI of 10$^\text{o} \times$10$^\text{o}$ around the Galactic center ($|b|<5^\text{o}$ and $|\ell|<5^\text{o}$) with $\Delta \Omega = 0.03$ steradians,
from Ref.~\cite{Cirelli:2010xx}.}
\end{center}
\end{table}

Taking into account all the aforementioned, in Fig.~\ref{sASTROGAMsensitivity} we show the e-ASTROGAM sensitivity for DM decay to a photon plus a second neutral particle ($N_{\gamma}^{(dec)} = 1$), considering the upper limits on $\langle \sigma v \rangle$ to a pair of photons presented by the e-ASTROGAM collaboration~\cite{eAstrogamAngelis:2017,Bartels:2017dpb} with $a=1/2$ and $N_{\gamma}^{(ann)}=2$. Several DM density profiles were used in the computation: NFW~\cite{Navarro:1995iw}, Moore~\cite{Diemand:2004wh}, Einasto~\cite{Graham:2005xx,Navarro:2008kc}, Einasto B~\cite{Tissera:2010} and Burkert~\cite{Burkert:1995yz}. For completeness, we show the functional forms of these profiles:
\begin{equation}
\begin{array}{r r l}
\text{NFW:} & \rho_{\text{NFW}}(r) = & \frac{\rho_{s}}{\frac{r}{r_{s}}\left(1+\frac{r}{r_{s}}\right)^{2}}, \\
\text{Moore:} & \rho_{\text{Moo}}(r) = & \frac{\rho_{s}}{\left(\frac{r}{r_{s}}\right)^{1.16}\left(1+\frac{r}{r_{s}}\right)^{1.84}}, \\
\text{Einasto:} & \rho_{\text{Ein}}(r) = & \rho_{s}\exp\left\{-\frac{2}{\alpha}\left[\left(\frac{r}{r_{s}}\right)^{\alpha}-1\right]\right\}, \\
\text{Burkert:} & \rho_{\text{Bur}}(r) = & \frac{\rho_{s}}{\left( 1+ \frac{r}{r_s} \right) \left(1+\left(\frac{r}{r_{s}}\right)^{2}\right)},
\end{array}
\label{profiles}
\end{equation}
where the Einasto (Einasto B) profile corresponds to $\alpha=0.17$ ($\alpha=0.11$). $r_s$ and $\rho_s$ represent typical scale radius and scale density, and the adopted values for each profile are shown in Table~\ref{DMprofilesTable}. These values satisfy $\rho_{\odot}= 0.3$ GeV/cm$^3$ at $r_{\odot}= 8.33$ kpc~\cite{Cirelli:2010xx}. 

The sensitivity of DM annihilation at 95\% CL derived by the e-ASTROGAM collaboration for DM annihilation~\cite{eAstrogamAngelis:2017,Bartels:2017dpb} considers a NFW profile, an effective observation time of one year, and a ROI of 10$^\text{o} \times$10$^\text{o}$ around the Galactic center ($|b|<5^\text{o}$ and $|\ell|<5^\text{o}$). An overview of the instrumental details can be found in Table 1 of Ref.~\cite{Bartels:2017dpb}.
We present the D-factors corresponding to that ROI in Table~\ref{DMprofilesTable}, for each DM profile~\cite{Cirelli:2010xx}. Since the only J-factor needed in our case is the one used to determine the annihilation projections, in 
Table~\ref{DMprofilesTable} we just show the NFW J-factor~\cite{Cirelli:2010xx}.

{To conclude, let us point out that we have followed a conservative approach in this computation, in the sense that we could have used a more efficient ROI for decay processes,
{such as a region of the Galactic halo, optimizing therefore the signal-to-background ratio, and as a consequence expecting a better performance.
For example, in the region of the Galactic halo with $60^\text{o}<|b|<90^\text{o}$ one has $\Delta\Omega = 1.68$ and a D-factor of 1.77 for a NFW profile, whereas for the ROI around the Galactic center we can see in Table~\ref{DMprofilesTable} that the D-factor is 11.8 but $\Delta\Omega = 0.03$. 
The ratio of the relevant product $\Delta\Omega$ multiplied by the D-factor gives 8.4.
With this naive argument we expect in principle a better performance using a ROI different from the one optimized for the annihilation analysis of e-ASTROGAM.
}

\bibliographystyle{utphys}
\bibliography{munussm}

\end{document}

%% file: defsV3.tex
\usepackage[T1]{fontenc}
\usepackage{epsfig}
\usepackage{latexsym}
\usepackage{graphicx}
\usepackage{amsmath}
\usepackage{amsfonts}   
\usepackage{amssymb}    
\usepackage{float}
\usepackage{bm}
\usepackage{url}
\usepackage{hyperref} 
\usepackage[nodisplayskipstretch]{setspace}
\def\lsim{\raise0.3ex\hbox{$\;<$\kern-0.75em\raise-1.1ex\hbox{$\sim\;$}}}
\def\gsim{\raise0.3ex\hbox{$\;>$\kern-0.75em\raise-1.1ex\hbox{$\sim\;$}}}

\newcommand{\captions}{\sf\caption}
\def    \beq            {\begin{equation}}
\def    \eeq            {\end{equation}}
\def    \bea           {\begin{eqnarray}}
\def    \eea           {\end{eqnarray}}

\def \mn{\mu\nu{\rm SSM}}

\def\g2{{\rm GeV}^2}

\def\sw2{sin^2 \theta_w}

\def\a^tau{\alpha_{\tau}}

\def\beq{\begin{equation}}
\def\eeq{\end{equation}}
\def\beqa{\begin{eqnarray}}
\def\eeqa{\end{eqnarray}}

\newcommand{\newc}{\newcommand}
\newc\BR{BR}
\newc{\akappa}{A_{\kappa} }
\newc\deltagmtwo{\delta (g-2)_{\mu}} 
\newc\deltaamu{\Delta a_{\mu}}

\def\anti{\overline}

\newc{\haa}{BR\(h_1\to a_1 a_1\)}
\newc{\abb}{BR\(a_1\to b\anti{b}\)}
\newc{\hbb}{BR\(h_1\to b\anti{b}\)}
\newc{\abund}{\Omega h^2}
\newc\bsgamma{b\rightarrow s \gamma }
\newc\bxsgamma{\overline{B}\rightarrow X_{s}\gamma}
\newc\brbsgamma{\BR(\overline{B}\rightarrow X_s\gamma)}

\newc{\Fermi}{\textit{Fermi}-}
